\documentclass[12pt]{iopart}                    

\usepackage{iopams}
\usepackage{latexsym,bbm}
\usepackage{graphicx,psfrag}
\usepackage{setspace}

\newcommand{\vp}{\varphi}
\newcommand{\mc}[1]{\mathcal{#1}}
\newcommand{\eps}{\varepsilon}
\renewcommand{\vec}[1]{\boldsymbol{#1}}
\newcommand{\tp}{\tilde{p}}
\newcommand{\TR}{\tilde{r}}
\newcommand{\tH}{\tilde{H}}
\newcommand{\lan}{\langle}
\newcommand{\ran}{\rangle}
\newcommand{\pa}{\partial}
\newcommand{\td}[1]{\tilde{#1}}

\begin{document}

\title[Next to leading order gravitational wave emission]{Next to leading order gravitational wave emission and dynamical
  evolution of compact binary systems with spin}

\author{D\"orte Hansen}
\address{Institute of Theoretical Physics,
         Friedrich-Schiller-University Jena,\\
         Max-Wien-Platz 1, D-07743 Jena}
\ead{D.Hansen@uni-jena.de}

\begin{abstract}
Compact binary systems with spinning components are considered. Finite
size effects due to rotational deformation are taken into
account. The dynamical evolution and next to leading order
gravitational wave forms are calculated, taking into account the orbital motion up to the first post-Newtonian approximation.
\end{abstract}

\submitto{\CQG}

\maketitle

\section{Introduction}
\label{sec:1}

Inspiralling compact binary systems are among the most promising
sources for the emission of gravitational waves detectable with
present day's gravitational wave interferometers. Earth bound
gravitational wave detectors such as Geo600, VIRGO, TAMA and LIGO 
are most sensitive at wavelengths of about 10-1000 Hz. This corresponds roughly to the last 10 minutes of the inspiral before final plounge. In that regime Newtonian mechanics is not valid and post-Newtonian approximation must be applied. An earlier stage of the inspiral process will be covered by the  yet to be built LISA interferometer. It is expected that LISA will be sensible to gravitational waves with frequencies from about $10^{-1}$ to $10^{-4}$ Hz. However, in order to actually detect gravitational waves highly accurate templates are essentially. Post-Newtonian corrections must be included into the EoM. Moreover, computing  the gravitational waveforms beyond the leading order approximation higher mass and current multipole moments must be taken into account.\\
In the past the study of close compact binaries was often based on the assumption that the stars can be treated as pointlike, non-spinning objects. Upon this assumption it is possible to derive an analytic, so called quasi-Keplerian  solution for the conservative part of the EoM up to the third post-Newtonian approximation \cite{Memmesheimer2004}.
Dissipative effects due to the emission of gravitational waves first appear at the order $(v/c)^5$, which corresponds to the 2.5 post-Newtonian order.\\
 In the presence of at least one spinning component a post-Newtonian spin-orbit coupling which first appears at 1.5 post-Newtonian order leads to complifications, which hamper the investigation of spinning compact binary systems enormeously. In general, neither orbital angular momentum nor the stellar spin are conserved. Up to now an analytic solution to the conservative part of the  EoM including spin-orbit coupling  has been found in a few special cases only (see e.g. \cite{Gopu}). 
\\
Thus far not much progress has been made in the investigation of close binary systems of finite size objects. Within the framework of post-Newtonian analysis it is often argued that 
finite size effects are neglegible during most of the inspiral process and will become important not until the last few orbits before the final plunge \cite{Wiseman1996}. Growing interest in the role of finite size effects comes mainly  from the side of numerical relativity. In fact, though finite size effects due to stellar rotation and oscillation in compact binary systems are very small they can well be in the order of the first post-Newtonian corrections to the orbital dynamics. These secular effects, being of Newtonian origin, accumulate over a large number of orbits and thus, seen at longer terms, lead to significant phase shifts in the gravitational waves. The influence of stellar oscillations on the dynamical evolution and leading order gravitational wave emission has been investigated by Kokkotas and Sch\"afer \cite{Kokkotas1995} and Lai and Ho \cite{Ho1999} for nonrotating, polytropic neutron stars and by Lai \emph{et al} \cite{Lai1994} and Hansen \cite{Hansen2005} for Riemann-S binaries. In all these approaches the analysis was based on Newtonian theory, the 2.5pN radiation reaction terms being the only post-Newtonian terms included.\\
This paper is devoted to the investigation of the influence of finite size effects on the dynamics and gravitational wave emission beyond the leading order approximation. Basically, perturbations of the point particle dynamics arise due to stellar oscillations (mainly tidally driven) and rotational deformation. Here we shall restrict ourselves to systems were the apsidal motion due to rotational deformation is much larger than the one caused by stellar oscillations.\\
To begin with, let us note that there is, strictly speaking, no spinning point particle. A spinning object automatically gains a finite size (see e.g. \cite{Schafer2004}). It hat been long known that the coupling of the nonvanishing stellar quadrupole moment to the orbital motion, which is a purely Newtonian effect, leads to an apsidal motion. For a couple of close main sequence star binaries this apsidal motion has been observed to great accuracy (see e.g. Claret and Willems \cite{Claret2002}). In these systems apsidal motion due to finite size effects is considerably larger than the relativistic periastron advance.
Another group of binary systems, where finite size effects play an important role, are binary pulsars such as PSR B 1259-63, which has been  found by Johnston \emph{et al.} in 1992 \cite{Johnston1992}.
While the compact component of PSR 1259-63 is a 47 ms pulsar, it's companion is a Be star, whose spin-induced quadrupole deformation leads to an apsidal motion.
\\
However, it is close compact binary systems which are most relevant for gravitational wave detectors. This includes not only black hole-black hole, black hole-neutron star or neutron star-neutron star binaries but, if LISA is ready to work, also close white dwarf binaries. Of course the rotational deformation of compact stars is much smaller than it would be possible in non-compact stars. However, at least for neutron star (NS) and white dwarf (WD) binaries one should not a priori neglect finite size effects due to quadrupole deformation.
In particular it is well possible that the perturbations introduced by the coupling of the stellar quadrupole moment to the orbital motion is of the same order of magnitude as the first post-Newtonian corrections. This is assumed throughout the paper. The analysis applies to compact binary systems, whose spinning component is a WD or a fast rotating NS. In order to  simplify calculations it is assumed that the spin is perpendicular to the orbital plane. In
section \ref{sec:2} the orbital evolution of a spinning compact binary is studied
up to first post Newtonian order in the point particle dynamics. The EoM as well as a parametric, quasi-Keplerian solution are
derived. 
In section \ref{sec:3} the next to leading order gravitational waveforms are
calculated explicitly. The long time evolution and the influence of
the quadrupole coupling to the inspiral process is discussed in
section \ref{sec:4}.

\section{The 1pN orbital motion including spin effects due to rotational deformation}
\label{sec:2}

In 1985 Damour and Deruelle \cite{Deruelle} succeeded in deriving an analytic solution to the 1pN EoM of a point-particle binary, which exhibits a remarkable similarity to the well known Kepler parametrization in Newtonian theory and is hence called quasi-Keplerian solution. Using the same strategy Wex considered a binary consisting of a pulsar and a spinning main sequence component at Newtonian order \cite{Wex}. Treating the Newtonian coupling between the rotationally deformed star and the orbital dynamics as a small perturbation he derived a quasi-Keplerian solution up to first order in the deformation parameter $q$, which will be introduced in the following. In this letter we shall extend his investigations, taking into account the orbital dynamics up to first post-Newtonian approximation. In order to derive an analytic solution we shall further assume that the modifications induced by finite size effects are of the same order as the 1pN orbital corrections. That is, we restrict our analysis to compact binary systems consisting of a fast spinning neutron star  or white dwarf  and a non-spinning compact object. Of course, all results can be applied to a mean sequence star-compact star binary in the Newtonian limit. \\
The rotational deformation of a spinning star of mass $m$ can be described by some parameter $q$, which is defined as \cite{Barker1970a}
\begin{equation}
     m q:=\frac{1}{2}\int dV' \rho(\vec r')\left[ r'^2-3(\hat{\vec s}\cdot\vec r')^2\right]
     =\Delta I, 
\end{equation}
where $m=\int \rho(\vec r')dV'$ is the stellar mass and $\Delta I$ is the difference of the moments of inertia parallel and perpendicular to the spin axis $\hat{\vec s}$, respectively. In particular, for rotating fluids, $q$ is given by (see e.g. \cite{Bildsten1995})
\begin{eqnarray}
     q=\frac{2}{3}kR^2\hat{\Omega}^2, \qquad \hat{\Omega}=\frac{\Omega}{\sqrt{Gm/R^3}}.
\end{eqnarray}
Here $R$ denotes the polar radius and $\Omega$ the angular velocity of the rotating star. The constant of apsidal motion $k$ strongly depends on the density distribution. It vanishes if all mass is concentrated in the center and takes it's maximal value $k_{max}=0.75$ for a homogeneous sphere. 
Neutron stars and low mass white dwarfs can be approximately modelled by an polytropic EoS with index n=$0.5\dots 1$ and $n\approx 1.5$ for neutron stars and white dwarfs, respectively. In fact, for polytropes the value of $k$ is given by  (see e.g. \cite{Cowling1938})
\begin{eqnarray*}
    k=\frac{1}{2}(\Delta_2(n)-1),
\end{eqnarray*}
where the function $\Delta_2(n)$ has been introduced by Chandrasekhar \cite{Chandrasekhar1933}.
\\
The quadrupole deformation  gives rise to a quadrupole coupling, which modifies the orbital dynamics as well as the gravitational wave emission of the binary. 
 Neglecting contributions arising from tidally induced stellar oscillations the quadrupole coupling can be described by a Hamiltonian
\begin{eqnarray}
\label{Hqgen}
    H_q=\frac{G\mc M\mu q}{2r^3}\left[\frac{3(\hat{\vec s}\cdot\vec r)^2}{r^2}-1\right],
\end{eqnarray}
where $r$ is the orbital separation, while $\mc M$ and $\mu$ denote total and reduced mass, respectively. As we shall see in the following this contribution will lead to a periastron advance already at Newtonian order.\\
In general, the direction of the star's spin won't be conserved, which complicates the analysis enormeously. However, one might assume scenarios where the spin is parallel to the angular momentum of the system, i.e. perpendicular to the orbital plane. In these cases Eq. (\ref{Hqgen}) takes a rather simple form,
\begin{eqnarray}
 \label{Hq}
   H_q=-\frac{G\mc M\mu q}{2r^3},
\end{eqnarray}
and we shall restrict to this special situation further on.\\
Before we proceed it is important to compare the perturbation introduced by quadrupole coupling with the first post-Newtonian correction terms to the EoM. Let us assume a spinning binary system in a circular orbit. Not taking into account the $q$-coupling the orbital energy reads\footnote{Note, that  $\mc E=H$.} 
\begin{eqnarray}
    \mc E_{orb} =-\frac{G\mc M\mu}{2r_0}+\frac{7-\nu}{8}\frac{G^2\mc M^2\mu}{r_0^2c^2}\equiv \mc E_N+\mc E_{pN}.
\end{eqnarray}
Comparing $\mc E_{pN}$ with the coupling energy we find that the $q$-term offers a contribution comparable to the 1pN orbital perturbation if
\begin{eqnarray*}
    \frac{q}{r_0}\approx \frac{G\mc M}{c^2} \equiv r_S.
\end{eqnarray*}
If $q/r_0$ is much larger than the Schwarzschild radius $r_S$ the Newtonian quadrupole contribution will clearly dominate, while for $q/r_0\ll G\mc M/c^2$ the leading perturbation to the Keplerian orbit comes from the post-Newtonian correction terms. The value of $q$ crucially depends on the density distribution (via $k$) and on the angular velorcity of the rotating star. The later one is bounded by the mass-shedding limit. For a Newtonian star with a polytropic EoS one can show that the mass-shedding limit is given by \cite{Teukolsky1983}
\begin{eqnarray}
    \Omega_{max}=\left(\frac{2}{3}\right)^{3/2}\sqrt{\frac{Gm}{R^3}},
\end{eqnarray}
where $R$ is the polar radius of the star. Thus, for polytropic stars $q$ is bounded by
\begin{eqnarray*}
    q_{max}=\frac{2}{3}kR^2\hat{\Omega}^2_{max}=\left(\frac{2}{3}\right)^4 kR^2.
\end{eqnarray*}
For a typical neutron star with polytropic index $n=1$ and polar radius $R=10$ km the maximal value of $q$ is $q_{max}=5.1$ km$^2$, while for a white dwarf with $n=1.5$  and a radius of 1000 km this value is given by\footnote{For $n=1$ the constant of apsidal motion is $k=0.259925$, for $n=1.5$ one finds $k=0.1446$.} $q_{max}=28563$ km$^2$. Now the post-Newtonian approximation is valid only outside the innermost stable circular orbit\footnote{In isotrope coordinates the radius of the last stable orbit of a test particle orbiting a Schwarzschild BH  is given by $r_{ISCO}= 5G\mc M/c^2$. For compact binaries a calculation of $r_{ISCO}$ in full General Relativity has not been successfull until now, but post-Newtonian analysis  suggests a value around $5G\mc M/c^2$.}.  For finite size binaries the orbital separation should be considerably larger than this value. To be more precisely, since we do not consider mass overflow the spinning star should remain well inside it's Roche volume throughout this calculation.
Taking this into account it becomes clear that for close NS-BH and NS-NS binaries the contribution of the quadrupole coupling is  by a factor 100 or more smaller than the first post-Newtonian correction. However, at least for sufficiently fast rotation the $q$-term gives rise to perturbations which are considerably larger than the 1.5 pN order. For spinning white dwarfs things can be different. In that case the finite size contribution due to rotational deformation can be equal or even larger than the pN contribution.
\\
These preliminary considerations showed that there might exist close compact binary systems for which the finite size effects introduced by rotational deformation are in the same range as the first post-Newtonian correction. Employing this assumption we shall now
 derive the quasi-Keplerian parametrization at the first post-Newtonian order including leading order quadrupole coupling (further on denoted as $q$-coupling).  Introducing the reduced energy and angular momentum, $E=:\mc E/\mu$ and $J=:\mc J/\mu$, respectively, the 1pN conserved energy including Newtonian $q$-coupling reads
\begin{equation}
 \label{energy}
\fl
    E=\frac{\vec v^2}{2}-\frac{G\mc M}{r}\left[1+\frac{q}{2r^2}\right] +\frac{1}{c^2}\left[
   \frac{3}{8}(1-3\nu)\vec v^4 +\frac{G\mc M}{2r}\left\{
   (3+\nu)\vec v^2+\nu\dot{r}^2 +\frac{G\mc M}{r}\right\} \right].
\end{equation}
Since the spin is parallel to the orbital angular momentum $\vec J$  the orbital plane is invariant in space. We are thus encouraged to introduce polar coordinates $r$ and $\vp$ in the usual way. Inserting $\vec v^2=\dot{r}^2+r^2\dot{\vp}^2$ into Eq. (\ref{energy}) one finds, after a little algebra, the 1pN exact EoM to be
\begin{eqnarray}
\label{1pNexact}
    \dot{\vp}&=\frac{J}{r^2}\left[1-\frac{1-3\nu}{c^2} E-\frac{G\mc M}{c^2r}(4-2\nu)\right],\\
    \dot{r}^2&=A+\frac{2B}{r}+\frac{C}{r^2}+\frac{D}{r^3},
\end{eqnarray}
where
\begin{eqnarray}
\label{coeffs}
    A&=2E\left[1+\frac{3}{2}(3\nu-1)\frac{E}{c^2}\right],\nonumber \\
    B&=G\mc M\left[1+(7\nu -6)\frac{E}{c^2}\right], \nonumber \\
    C&=-J^2+\frac{1}{c^2}\left[ 2(1-3\nu)EJ^2+(5\nu-10)G^2\mc M^2\right],\nonumber \\
    D&=G\mc Mq+(8-3\nu)\frac{G\mc MJ^2}{c^2}
\end{eqnarray}
are constants. In the standard approach of Damour and Deruelle it is crucial that $D$ is of order $O(c^{-2})$ and thus a small quantity. Now in our case $D$ depends not only on $c$ but it is also linear in the deformation parameter $q$. If the spinning component is governed by a soft EoS the correction to the orbital motion induced by the $q$-coupling is much larger than the 1pN corrections. This is usual the case for main-sequence star binaries (see e.g. Claret and Willems \cite{Claret2002}). For compact stars in close binary systems, on the other hand, the contribution of the $q$-coupling can be of the same order as the 1pN orbital correction. Under this assumption we can apply Damour and Deruelles strategy straightforwardly, deriving a quasi-Keplerian solution up to linear order of $q$. This yields
\begin{eqnarray}
\label{sol}
    r&=a_r(1-e_r\cos u),\qquad 
    u-e_t\sin u=n(t-t_0), \\
\label{sol2}
    \vp&=2(\kappa+1)\mathrm{arctan}\left[\sqrt{\frac{1+e_\vp}{1-e_\vp}}\tan\frac{u}{2}\right].
\end{eqnarray}
where $n, a_r, e_r, e_t$ and $e_\vp$ depend on the coefficients defined above as 
\begin{eqnarray}
  a_r&=-\frac{B}{A}-\frac{C}{2J^2}, \qquad
  e_\vp= e_t\left[1-\frac{A}{B}\left( 2(\nu-2)\frac{G\mc M}{c^2} +\frac{D}{J^2}\right)
         \right], \\
  e_r&=e_t\left[1-\frac{DA}{2BJ^2}\right],
  \hspace{1.5cm}
  e_t=\sqrt{1-\frac{A}{B^2}\left[C+\frac{BD}{J^2}\right]} 
\end{eqnarray}
and $n=\sqrt{-A^3/B^2}$, and $\kappa$ is given by
\begin{eqnarray}
     \kappa=\frac{3G^2\mc M^2}{J^2}\left[\frac{1}{c^2}+\frac{q}{2J^2}\right].
\end{eqnarray}
As it turns out, the parameter which has to be small for this solution to hold is $\delta\equiv q/J^2 \propto 1/c^2$. Expressing the parameters of the quasi-Keplerian solution in terms of the 1pN conserved energy and $\delta$, we find
\begin{eqnarray*}
    a_r&=-\frac{G\mc M}{2E}-\frac{1}{2}G\mc M\delta +\frac{G\mc M}{4c^2}(\nu -7),\\
    n^2&= -\frac{8E^3}{G^2\mc M^2}\left[1-\frac{\nu-15}{2}\frac{E}{c^2}\right] \\
     e_t&=e_r\left[1+E\left\{\delta+\frac{8-3\nu}{c^2}\right\}\right],\\
     e_\vp&=e_r\left[1-E\left\{\delta+\frac{\nu}{c^2}\right\}\right].
\end{eqnarray*}

\subsection{Hamiltonian formulation}
\label{sec:2a}

The quasi-Keplerian solution given above describes the dynamics of a spinning compact binary with Newtonian quadrupole coupling at the first post-Newtonian order. However, according to GR the system looses energy due to the emission of gravitational waves, beginning at the order $(v/c)^5$ in post-Newtonian approximation schemes. There are basically two ways to study the dynamical evolution of the binary system. In one approach the gravitational wave emission is considered as a secular effect and the dissipative terms do not enter the EoM. This allows the derivation of the quasi-Keplerian parametrization given above. Now we shall follow the other approach, where the radiation reaction terms are included into the EoM. In a first step we derive the Hamiltonian formulation to the conservative system. Energy loss due to emission of gravitational waves is incorporated by a time dependent radiation reaction Hamiltonian in a second step.
\\
The total Hamiltonian of a spinning compact binary system up to first post-Newtonian approximation reads
\begin{eqnarray}
    H_{1pN}=H_{N}+H_{q}+H_{pN},
\end{eqnarray}
where $H_N$ and $H_{pN}$ denote the point-particle Hamiltonian at Newtonian and first post-Newtonian order, respectively. If we would extend our analysis up to 2pN, other spin-depended terms would be present: at 1.5pN order the relativistic spin-orbit coupling enters into the Hamiltonian. It is, among others, responsible for the Lense-Thirring effect. At the level of the second post-Newtonian approximation the relativistic spin-spin coupling leads to a precession of the orbital plane. Both, incorporating spin-orbit as well as investigating the spin-spin coupling, is beyond this letter, which is devoted to the study of the influence of certain finite size effects on the dynamical evolution and gravitational wave emission of the binary system.\\
In the center of mass system the 1pN Hamiltonian, including $q$-coupling, reads
\begin{eqnarray}
\label{Htot}
\fl
     H_{1pN}=\frac{1}{2\mu}\left[p_r^2+\frac{p_\vp^2}{r^2}\right]-
     \frac{G\mc M\mu}{r} -\frac{G\mc M\mu }{2r^3}q 
     +\frac{1}{c^2}\left[\frac{3\nu-1}{8\mu^3}\left\{p_r^4+2p_r^2\frac{p_\vp^2}{r^2}+\frac{p_\vp^4}{r^4}\right\} 
    \right. \nonumber \\
     \left. 
      -\frac{G\mc M}{2r}\left\{(3+2\nu)\frac{p_r^2}{\mu}+(3+\nu)\frac{p_\vp^2}{\mu r^2}\right\} +\frac{G^2\mc M^2\mu}{2r^2}\right].
\end{eqnarray}
Note, that $H_{1pN}=\mc E=\mu E$, since $H_{1pN}$ is conserved at the first post-Newtonian order. As  has been already mentioned before,
 the quadrupole interaction term is de facto a \emph{Newtonian} correction to the point particle Hamiltonian. This is important to keep in mind when, for instance, calculating the orbital evolution and gravitational wave emission of binary pulsars with a main sequence star companion, such as PSR B1259-63 \cite{Wex}. For the Hamiltonian equations that govern the time evolution of the binary system one finds
\begin{eqnarray}
 \label{hameom}
  \dot{r}&=\frac{p_r}{\mu}\left[
   1+\frac{1}{\mu^2c^2}\left\{
   \frac{3\nu-1}{2}\left(p_r^2+\frac{p_\vp^2}{r^2}\right)-\frac{G\mc M\mu^2}{r}(3+2\nu)\right\}
   \right],\\
  \dot{\vp}&= \frac{p_\vp}{\mu r^2}\left[
   1+\frac{1}{\mu^2c^2}\left\{
   \frac{3\nu-1}{2}\left(p_r^2+\frac{p_\vp^2}{r^2}\right)-\frac{G\mc M\mu^2}{r}(3+\nu)
  \right\}\right],\nonumber \\
  \dot{p}_r&=\frac{p_\vp^2}{\mu r^3}-\frac{G\mc M\mu}{r^2}\left(1+\frac{3q}{2r^2}\right) +\frac{1}{ c^2}\left[
   \frac{3\nu-1}{2\mu^3}\frac{p_\vp^2}{r^3}\left(p_r^2+\frac{p_\vp^2}{r^2}\right) \right.
   \nonumber \\
 &\hspace{0.5cm}
   \left. 
   +\frac{G^2\mc M^2\mu}{r^3}-\frac{G\mc M}{2\mu r^2}\left\{ (3+2\nu)p_r^2 +3(3+\nu)\frac{p_\vp^2}{r^2}
   \right\} \right],
    \nonumber \\
   \dot{p}_\vp&=0. \nonumber
\end{eqnarray}
Now let us include leading order dissipative effects into that scheme. In general, the leading order energy dissipation of a matter distribution is governed by the time-dependent radiation reaction Hamiltonian \cite{Schafer1990}
\begin{eqnarray}
    H_{reac}(t)=\frac{2G}{5 c^5}\mc I^{(3)}_{ij}(t)\int dV \left[\frac{\pi_i\pi_j}{\rho} 
    +\frac{1}{4\pi G}\pa_i U \pa_j U\right],
\end{eqnarray}
where $\mc I_{ij}$ is the symmetric tracefree mass quadrupole tensor of the matter distribution, $\rho$ is the coordinate rest-mass density, $\pi_i$ the momentum density and $U$ is the gravitational potential that satisfies the Poisson equation with source term $\rho$. Since we are treating the $q$-dependent terms as being formally of first post-Newtonian order, the components of the binary can be considered as pointlike objects troughout the calculation of $H_{reac}(t)$. Thus in the center of mass frame the radiation reaction Hamiltonian describing the leading order energy dissipation due to gravitational wave emission is given by
\begin{eqnarray}
 \label{Hreac}
  H_{reac}(t)=\frac{2G}{5c^5}\mc I^{(3)}_{ij}(t)\left[\frac{p_ip_j}{\mu}-G\mc M\mu\frac{x^ix^j}{r^3}\right].
\end{eqnarray}
It is crucial to consider $\mc I^{(3)}_{ij}(t)$ as a function of time, and not as a function of generalized coordinates and momenta, when calculating the radiation reaction part of the EoM according to
\begin{eqnarray*}
   (\dot{p}_i)_{reac}=-\frac{\pa H_{reac}}{\pa q^i},\qquad
   (\dot{q}^i)_{reac}=\frac{\pa H_{reac}}{\pa p_i}.
\end{eqnarray*}
Only afterwards $\mc I^{(3)}_{ij}$ can be expressed as a function of $p_r,p_\vp,r$ and $\vp$. Explicitly, the calculation yields \cite{Kokkotas1995},\cite{Hansen2005}
\begin{eqnarray}
 \label{radterms}
  (\dot{p}_r)_{rad}&=\frac{8}{3}\frac{G^2p_r}{r^4c^5}\left(\frac{G\mc M^3\nu}{5}-\frac{p_\vp^2}{\nu r}\right), \nonumber \\
  (\dot{p}_\vp)_{rad}&=-\frac{8}{5}\frac{G^2p_\vp}{\nu r^3c^5}\left(\frac{2G\mc M^3\nu^2}{r}+2\frac{p_\vp^2}{r^2}-p_r^2\right), \nonumber \\
  (\dot{r})_{rad}&=-\frac{8}{15}\frac{G^2}{\nu r^2c^5}\left(2p_r^2+6\frac{p_\vp^2}{r^2}\right), \nonumber \\
  (\dot{\vp})_{rad}&=-\frac{8}{3}\frac{G^2p_rp_\vp}{\nu r^4c^5}.
\end{eqnarray}
For numerical calculations it is useful to introduce scaled variables such that $G=c=1$. The corresponding scaling is given by
\begin{eqnarray*}
    p_r=\mu c\tp_r,\quad
    p_\vp=\frac{G\mc M\mu}{c}\tp_\vp,\quad
    r=\frac{G\mc M}{c^2}\TR,\quad
    H=\mu c^2\tH,\quad
    q=\frac{G^2\mc M^2}{c^4}\tilde{q}.
\end{eqnarray*}
Applying this, the Hamiltonian equations governing the evolution of the binary system including leading order radiation back reaction read
\begin{eqnarray}
\fl
\label{eomscal1}
    \dot{\TR}=\tp_r\left[ 1+\frac{3\nu-1}{2}\left\{\tp_r^2+\frac{\tp_\vp^2}{\TR^2}\right\} -
    \frac{3+2\nu}{\TR} \right]
    -\frac{8}{15}\frac{\nu}{\TR^2}\left[2\tp_r^2+6\frac{\tp_\vp^2}{\TR^2}\right], 
    \\
\fl
\label{eomscal2}
   \dot{\vp}=\frac{\tp_\vp}{\TR^2}\left[1+\frac{3\nu-1}{2}\left\{
    \tp_r^2+\frac{\tp_\vp^2}{\TR^2}\right\} -\frac{3+\nu}{\TR}\right] -\frac{8\nu}{3}\frac{\tp_r\tp_\vp}{\TR^4},  \\
\fl
\label{eomscal3}
    \dot{\tp}_r=\frac{\tp_\vp^2}{\TR^3}-\frac{1}{\TR^2}\left[1+\frac{3\tilde{q}}{2\TR^2}\right]
    +\frac{3\nu-1}{2}\frac{\tp_\vp^2}{\TR^3}\left[\tp_r^2+\frac{\tp_\vp^2}{\TR^2}\right]
    -\frac{3+2\nu}{2\TR^2}\tp_r^2 -\frac{3}{2}(3+\nu)\frac{\tp_\vp^2}{\TR^4} \nonumber \\
    +\frac{1}{\TR^3} +\frac{8\nu}{3}\frac{\tp_r}{\TR^4}\left[\frac{1}{5}-\frac{\tp_\vp^2}{\TR}\right],
     \\
\fl
\label{eomscal4}
    \dot{\tp}_\vp=-\frac{8\nu}{5}\frac{\tp_\vp}{\TR^3}\left[\frac{2}{\TR} +2\frac{\tp_\vp^2}{\TR^2}
    -\tp_r^2\right].
\end{eqnarray}
Neither the total energy nor the orbital angular momentum is conserved, as indicated by Eq. (\ref{eomscal4}).\\
The time evolution of binary systems described by Eqs. (\ref{eomscal1})-(\ref{eomscal4}) is fully determined by 3 parameters: the semi-major axis $a_r$, the orbital eccentricity $e_r$
and the deformation parameter $q$, which have to be known for $t=0$. Starting the numerical integration in the periastron, i.e. at $\vp(0)=0$, the initial values for $r$ and $p_r$ follow immediately as
\begin{eqnarray*}
    r(0)=r_0=a_r(0)(1-e_r(0)),\qquad
    p_r(0)=0.
\end{eqnarray*}
To determine the initial value for $p_\vp$ we use that, at the beginning of the integration, the total energy of the system is given by the conservative part of the Hamiltonian, or, using the reduced energy, $E(0)=E=H_{1pN}/\mu$. It follows then from Eq. (\ref{Htot}) that\footnote{Note that $p_\vp \to \mu p_\vp$ and $p_r \to \mu p_r$.}
\begin{eqnarray*}
   p_\vp(0)^2&=2r_0^2E_{1pN}+2G\mc Mr_0\left[1+\frac{q}{2r_0^2}\right]+\frac{1}{c^2}\left[
   (1-3\nu)r_0^2(E_N^{orb})^2
   \right. \\
   &\hspace{0.5cm}
   \left. 
   +4(1-\nu)G\mc Mr_0E_N^{orb} 
   +(6-\nu)G^2\mc M^2\right],
\end{eqnarray*}
where $E_N^{orb}$ is the Newtonian energy of the orbit.

\section{Higher order gravitational wave emission}
\label{sec:3}

In the previous sections we have investigated the dynamical evolution of a spinning compact binary system at the first post Newtonian approximation. Now we shall turn our attention to the gravitational waves emitted by the system. Far away from the source the space time can be assumed to be asymptotically flat, such that the metric is locally Minkowskian.  In fact, in  asymptotically flat space-times the gravitational waves emitted by an isolated binary systems are expected to obey a multipole expansion of the form (see e.g. \cite{Thorne1980})
\begin{eqnarray}
\label{rad1}
    h^{TT}_{ij}&=\frac{G}{Dc^4}P_{ijkm}(\vec N)\sum_{l=2}^\infty 
    \left[\left(\frac{1}{c}\right)^{l-2}\left(\frac{4}{l!}\right)\mc I^{(l)}_{kmA_{l-2}}(t-D/c)N_{A_{l-2}} \right. \nonumber \\
    &\hspace{0.5cm}
    \left. 
     +\left(\frac{1}{c}\right)^{l-1}
    \left(\frac{8l}{(l+1)!}\right)
     \eps_{pq(k}\mc J^{(l)}_{m) pA_{l-2}}(t-D/c)N_qN_{A_{l-2}}\right],
\end{eqnarray}
where $h_{ij}^{TT}$ is the symmetric-tracefree (STF) part of the metric perturbation $h_{ij}$, the brackets denote symmetrization and $D$ is the source-observer distance. The unit vector $\vec N$   points from the binary to the observer and $A_l=a_1a_2\dots a_l$  ($a_i=1,2,3$)  is a multi-index. $\mc I_{A_l}$ and $\mc J_{A_l}$ are the STF mass and current multipole moments that parameterize the radiation field in a Cartesian coordinate frame. However, if the direction of the angular momentum is conserved, it is more suitable to use STF-multipole moments $I^{lm}$ and $S^{lm}$, $m=-l,\dots,l$, that are irreducibly defined with respect to the axis of angular momentum. The relation of these to the Cartesian multipole components are given in Eqs. (\ref{rel1}) and (\ref{rel2}) in the appendix. From Eq. (\ref{rad1}) it is then derived that the radiation field $h_{ij}^{TT}$, expressed in terms of time derivatives of $I^{lm}$ and $S^{lm}$, is given by
\begin{eqnarray}
\fl
\label{rad2}
    h_{ij}^{TT}=\frac{G}{Dc^4}\sum_{l=2}^\infty \sum_{m=-l}^l
   \left[
   \left(\frac{1}{c}\right)^{l-2}I^{(l)lm}(t-D/c)T^{E2,lm}_{ij}(\Theta,\Phi) 
  \right. \nonumber \\
   \left.
   +\left(\frac{1}{c}\right)^{l-1}S^{(l)lm}(t-D/c)T^{B2,lm}_{ij}(\Theta,\Phi)\right],
\end{eqnarray}
where $T^{E2,lm}_{ij}$ and $T^{B2,lm}_{ij}$ are the so-called pure-spin tensor-spherical harmonics of electric and magnetic type. These harmonics are orthonormal on the unit sphere. In fact, introducing unit vectors $\hat{\Phi}$ and $\hat{\Theta}$, they can be decomposed into a term proportional to $(\hat{\Theta}\otimes\hat{\Theta}-\hat{\Phi}\otimes\hat{\Phi})$ and $(\hat{\Theta}\otimes\hat{\Phi}+\hat{\Phi}\otimes\hat{\Theta})$, respectively. That way, if the $T^{E/B2,lm}$ are known, one obtains the polarization states $h_+$ and $h_\times$ of the radiation field from Eq. (\ref{rad2}) without any further calculations. The pure-spin tensor-spherical harmonics needed here are given by Eqs. (\ref{pure1})-(\ref{pure10}) in the appendix\footnote{In this representation, $h_+$ is the $(\hat{\Theta}\otimes\hat{\Theta}-\hat{\Phi}\otimes\hat{\Phi})$-part of Eq. (\ref{rad2}), while $h_\times$ is the $(\hat{\Theta}\otimes\hat{\Phi}+\hat{\Phi}\otimes\hat{\Theta})$-part.}. \\
In section 2 the point particle contribution was taken into account up to first post-Newtonian approximation, and the quadrupole coupling term, though present already at Newtonian order, was assumed to be of the same order as the 1pN corrections to the point particle dynamics. That means, we have to go beyond the leading order gravitational wave formula. Considering the dynamics up to 1pN requires the application of the multipole expansion (\ref{rad2}) up to $l=4$ for the mass multipole moments and up to $l=3$ for the current multipole moments. Explicitly, neglecting all higher order terms, Eq. (\ref{rad2}) is reduced to\footnote{From now on we omit the supscript $TT$ for notational convenience.}
\begin{eqnarray}
\label{rad3}
\fl
     h_{+,\times}=\frac{G}{Dc^4} \left[ 
    \sum_{m=-2}^2 I^{(2)2m}T^{E2,2m}_{+,\times} +\frac{1}{c}\left\{
    \sum_{m=-3}^3 I^{(3)3m}T^{E2,3m}_{+,\times} 
    +\sum_{m=-2}^2 
     S^{(2)2m}T^{B2,2m}_{+,\times}\right\} \right. \nonumber \\
    \left.
    +\frac{1}{c^2}\left\{
    \sum_{m=-4}^4I^{(4)4m}T^{E2,4m}_{+,\times}+\sum_{m=-3}^3S^{(3)3m}T^{B2,3m}_{+,\times}
   \right\} \right] 
\end{eqnarray}
or, in a more convenient form, 
\begin{eqnarray}
  h_{+,\times}=h_{+,\times}^{(0)}+\frac{1}{c}h_{+,\times}^{(1)}+\frac{1}{c^2}h_{+,\times}^{(2)}.
\end{eqnarray}
Note that according to Eq. (\ref{rad3}) except for $I^{(2)2m}$ all other time derivatives and multipole moments are required only at leading order.
\\
Let us start by noting that, in the center-of-mass system, the mass and current multipole moments of a two-body system are given by 
\begin{eqnarray}
 \label{multi}
  \mc I_{ij}&=\mu x_{\lan i}x_{j\ran}\left[1+\frac{29}{42}(1-3\nu)\frac{v^2}{c^2}-\frac{5-8\nu}{7c^2}\frac{G\mc M}{r}\right] \nonumber \\
  & 
  +\frac{\mu(1-3\nu)}{21 c^2}\Bigl[ -12(\vec r\cdot\vec v)x_{\lan i}v_{j\ran} +11 r^2 v_{\lan i}v_{j\ran}\Bigr] +\mc I_{ij}^s,\\
  \mc I_{ijk}&=-\mu\sqrt{1-4\nu} \,x_{\langle i}x_jx_{k\rangle}, \\
  \mc I_{ijkl}&=\mu(1-3\nu)x_{\lan i}x_jx_kx_{l\ran},\\
  \mc J_{ij}&=-\mu\sqrt{1-4\nu}\eps_{ab\langle i}x_{j\rangle}x_av_b,\\
  \mc J_{ijk}&=\mu(1-3\nu)\eps_{ab\lan k}x_ix_{j\ran}x_av_b,
\end{eqnarray}
where brackets denote the STF-part of the correspondent tensor (see appendix A).
At this point it is neccessary to consider a moment the contribution of the stellar mass-quadrupole moment. Since one component of the binary is spinning and thus automatically gains a finite size there could be, in principle, a contribution of $\mc I_{ij}^{s}$ to the gravitational wave emission of the system. This contribution is, however, very small, unless the energy stored in the internal stellar degrees of freedom, e.g. oscillations of the star, is comparable to the orbital energy. 
From now on we shall assume that $\ddot{\mc I}^{s}_{ij}$ is either trivial or can be neglected compared to all other terms present in the calculation. \\
Using the Newtonian equation of motion for a point-particle binary\footnote{Remember that $q$ is treated formally as a 1pN quantity.}
\begin{equation*}
    \dot{\vec v}=-\frac{G\mc M}{r^3}\vec r
\end{equation*}
to calculate $\mc I^{(3)3m}, \mc I^{(4)4m}, \mc J^{(2)2m}$ and $\mc J^{(3)3m}$ and 
the 1pN equation of motion in the form
\begin{eqnarray*}
\fl
    \dot{\vec v}
    =-\frac{G\mc M}{r^3}\vec r\left(1+\frac{3q}{2r^2}\right) 
    +\frac{G\mc M}{r^3c^2}\left[\vec r\left\{\frac{G\mc M}{r}(4+2\nu)-v^2(1+3\nu)
    +\frac{3\nu}{2}\frac{(\vec r\cdot\vec v)^2}{r^2}\right\} \right. \\
    \Bigl.
     +(4-2\nu)(\vec r\cdot\vec v)\vec v\Bigr] \nonumber
\end{eqnarray*}
to calculate $\mc I^{(2)}_{ij}$ one finds, after some lengthy calculations (see also \cite{Junker1992})
\begin{eqnarray}
 \label{massquadru}
\fl
  \mc I_{ij}^{(2)}=2\mu v_{\lan i}v_{j\ran}\left[1+\frac{9}{14}(1-3\nu)\frac{v^2}{c^2}+\frac{54\nu-25}{21}\frac{G\mc M}{r c^2}\right]
   \nonumber \\
  +2\mu x_{\lan i}v_{j\ran}\frac{25+9\nu}{7}\frac{G\mc M}{r^3c^2}(\vec r\cdot\vec v) 
  -\mu x_{\lan i}x_{j\ran}\frac{G\mc M}{r^3}\left[2\left(1+\frac{3q}{2r^2}\right)+\frac{61+48\nu}{21}\frac{v^2}{c^2}
   \right. \nonumber \\
   \left.
   -\frac{2}{7c^2}(1-3\nu)\frac{(\vec r\cdot\vec v)^2}{r^2}-(10-\nu)\frac{G\mc M}{r c^2}\right], \\
\label{massoktu}
\fl
   \mc I_{ijk}^{(3)}=-\mu\sqrt{1-4\nu}\left[9\frac{G\mc M}{r^5}(\vec r\cdot\vec v)x_{\langle i}x_jx_{k\rangle} -21\frac{G\mc M}{r^3}v_{\langle i}x_jx_{k\rangle} +6v_{\langle i}v_jv_{k\rangle}\right],\\
\fl
\label{masshexa}
  \mc I^{(4)}_{ijkl}=4\mu(1-3\nu)\left[6v_{\lan i}v_jv_kv_{l\ran}-48\frac{G\mc M}{r^3}v_{\lan i}v_jx_kx_{l\ran} +42\frac{G\mc M}{r^5}(\vec r\cdot\vec v)v_{\lan i}x_jx_kx_{l\ran} \right. \nonumber \\
    \left.
    +\frac{G\mc M}{r^3}x_{\lan i}x_jx_kx_{l\ran}\left\{7\frac{G\mc M}{r^3}+3\frac{v^2}{r^2}-15\frac{(\vec r\cdot\vec v)^2}{r^4}\right\}\right],
\end{eqnarray}
while the time derivatives of the current multipole moments read
\begin{eqnarray}
\label{currentquadru}
    \mc J_{ij}^{(2)}&=\mu\sqrt{1-4\nu}\frac{G\mc M}{r^3}\eps_{ab\langle i}x_{j\rangle}x_av_b,\\
 \label{currentoctu}
  \mc J^{(3)}_{ijk}&=2\mu\frac{G\mc M}{r^3}(1-3\nu)\left[-4\eps_{ab\lan k}x_iv_{j\ran}x_av_b+3\frac{\vec r\cdot\vec v}{r^2}\eps_{ab\lan k}x_ix_{j\ran}x_av_b\right].
\end{eqnarray}
In particular, within this approximation the $q$ coupling is only relevant for the second time derivative of the mass quadrupole tensor. Using the relation between the two classes of multipole moments given by Eqs. (\ref{rel1}) and (\ref{rel2}) we get the following expressions for the time derivatives of $I^{lm}$ and $S^{lm}$, now in polar coordinates:
\begin{eqnarray}
\fl
    I^{(2)20}=\frac{4\mu}{3}\sqrt{\frac{3\pi}{5}} \left[
    -2\left\{v^2-\frac{G\mc M}{r}\left(1+\frac{3q}{2r^2}\right)
    \right\} +\frac{1}{c^2}\left\{
    \frac{G^2\mc M^2}{r^2}(\nu -10)
    \right. \right. \nonumber \\
    \left. \left. 
    +\frac{9}{7}(3\nu -1)v^4 
    +\frac{G\mc M}{7r}( (37-20\nu)r^2\dot{\vp}^2 -(15+32\nu)\dot{r}^2)
    \right\} \right], 
\end{eqnarray}
\begin{eqnarray}
\fl
    I^{(2)21}=0, \\
\fl
    I^{(2)22}=\sqrt{\frac{8\pi}{5}} \mu e^{-2i\vp} \left[
    2\left\{ 
    \dot{r}^2-r^2\dot{\vp}^2 -\frac{G\mc M}{r} \left(1+\frac{3q}{2r^2}\right)
    -2ir\dot{r}\dot{\vp} \right\} \right.  \\
    +\frac{1}{c^2}\left\{
     (10-\nu)\frac{G^2\mc M^2}{r^2} 
     +\frac{G\mc M}{21 r} ( 3(15+32\nu)\dot{r}^2 -(11+156\nu)r^2\dot{\vp}^2) 
     \right. \nonumber \\
     \left. 
     \left. 
     +\frac{9}{7}(1-3\nu)(\dot{r}^4-r^4\dot{\vp}^4) 
     -ir\dot{r}\dot{\vp} \left( \frac{10}{21}(5+27\nu)\frac{G\mc M}{r}
     +\frac{18}{7}(1-3\nu)v^2\right) \right\}\right],  \nonumber
\end{eqnarray}
\begin{eqnarray}
\fl
     I^{(3)30}=0,\\
\fl
     I^{(3)31}=4\nu(m_1-m_2)\sqrt{\frac{\pi}{35}}e^{-i\vp} \left[ \dot{r}\left\{2\frac{G\mc M}{r}-v^2\right\} +ir\dot{\vp}\left\{ v^2-\frac{7}{6}\frac{G\mc M}{r}
     \right\}\right], \\
\fl
     I^{(3)33}=2\nu(m_1-m_2)\sqrt{\frac{\pi}{21}} e^{-3i\vp} \left[2 \left\{
     \dot{r}^2-2\frac{G\mc M}{r}-3r^2\dot{\vp}^2\right\} \dot{r} 
     \right. \nonumber \\
     \left. 
     +ir\dot{\vp} \left\{ 7\frac{G\mc M}{r} -6 \dot{r}^2+2r^2\dot{\vp}^2 \right\}\right], \\
\fl 
     I^{(4)40}=\frac{2}{21}(1-3\nu)\mu\sqrt{\frac{\pi}{5}} \left[7\frac{G^2\mc M^2}{r^2}
     -\frac{G\mc M}{r}\left\{ 18\dot{r}^2+13r^2\dot{\vp}^2\right\} +6v^4\right], \\
\fl
     I^{(4)41}=I^{(4)43}=0,\\
\fl 
     I^{(4)42}=\frac{2}{63}\sqrt{2\pi} (1-3\nu) \mu e^{-2i\vp}\left[
     -7\frac{G^2\mc M^2}{r^2}+\frac{G\mc M}{r}(18\dot{r}^2-3r^2\dot{\vp}^2)
     \right. \nonumber \\  
     \left. 
     +6r^4\dot{\vp}^4 -6\dot{r}^4  
     +3ir\dot{r}\dot{\vp}\left\{ 4v^2 -9\frac{G\mc M}{r}\right\} \right],\\
\fl  
    I^{(4)44}=\frac{2}{9}\sqrt{\frac{\pi}{14}}(1-3\nu)\mu e^{-4i\vp} \left[
     7\frac{G^2\mc M^2}{r^2} +\frac{G\mc M}{r} (-18\dot{r}^2+51r^2\dot{\vp}^2) +6\dot{r}^4
     \right. \nonumber \\
    \left.
    - 36r^2\dot{r}^2\dot{\vp}^2 +6r^4\dot{\vp}^4 
     +ir\dot{r}\dot{\vp} \left\{54\frac{G\mc M}{r}-24\dot{r}^2+24r^2\dot{\vp}^2\right\}\right],\\
\fl   
     S^{(2)20}=S^{(2)22}=0,\\
\fl 
     S^{(2)21}=\frac{8}{3}\sqrt{\frac{2\pi}{5}}(m_1-m_2) G\mc M\nu \dot{\vp}e^{-i\vp}, \\
\fl 
     S^{(3)30}=-4\sqrt{\frac{\pi}{105}}(1-3\nu)G\mc M\mu \dot{r}\dot{\vp}, \\
\fl  
     S^{(3)31}=S^{(3)33}=0,\\
\fl 
     S^{(3)32}=\frac{2}{3}\sqrt{\frac{2\pi}{7}}(1-3\nu)G\mc M\mu e^{-2i\vp} \dot{\vp}(\dot{r}-4ir\dot{\vp}).
\end{eqnarray}
Here we used that $\sqrt{1-4\nu}=(m_1-m_2)/\mc M$. Since only $S^{(2)21},\ I^{(3)31}$ and $I^{(3)33}$ contribute to the first correction to the leading order quadrupole formula $h_{+,\times}^{(1)}$, it becomes clear  that $h_{+,\times}^{(1)}$ vanishes for equal mass binary systems. In that case the first nontrivial correction to $h_{+,\times}^{(0)}$ is of order $1/c^2$. \\
It is possible to derive, after some straightforward but rather lengthy calculations, analytic expressions for the time derivatives of the multipole moments in terms of $e_r, E$ and $\delta$; the corresponding relations are given in appendix C. For our purposes it is, however, more useful to express the polarization states of the gravitational radiation field in terms of generalized coordinates and velocities or -- in Hamiltonian formulation -- in terms of generalized coordinates and momenta. Inserting above relations in Eq. (\ref{rad3}) one finds, at leading order,
\begin{eqnarray}
\fl 
     \frac{Dc^4}{G}h_+^{(0)}=(1+\cos^2\Theta)\mu\left[ \cos 2(\Phi-\vp)\left\{
     \dot{r}^2-r^2\dot{\vp}^2-\frac{G\mc M}{r}\left(1+\frac{3q}{2r^2}\right)\right\}
     \right. \nonumber \\
     \Bigl. 
     +2r\dot{r}\dot{\vp}\sin 2(\Phi-\vp)\Bigr] 
     -\mu \sin^2\Theta\left[\dot{r}^2+r^2\dot{\vp}^2-\frac{G\mc M}{r}\left(
     1+\frac{3q}{2r^2}\right)\right],  \\
\fl 
     \frac{Dc^4}{G}h_\times^{(0)}=\mu\cos\Theta\left[
     4r\dot{r}\dot{\vp}\cos 2(\Phi-\vp) -2\sin 2(\Phi-\vp) \left\{
     \dot{r}^2-r^2\dot{\vp}^2
     -\frac{G\mc M}{r}\left(1+\frac{3q}{2r^2}\right)\right\}\right]. \nonumber 
\end{eqnarray}
Note that, in order to emphasize the character of the $q$-coupling, the $q$-dependent terms have been included into the leading order component of the radiation field. Defining $\Delta m\equiv m_1-m_2$ the first correction terms read
\begin{eqnarray}
\fl 
    \frac{Dc^5}{G}h_{+}^{(1)}=\frac{\Delta m}{\mc M}\mu\sin\Theta \left[\frac{4}{3}G\mc M\dot{\vp}\sin(\Phi-\vp) 
    +\frac{3\cos^2\Theta-1}{2}\left\{ \left(2\frac{G\mc M}{r}-v^2\right)
    \dot{r}\cos(\Phi-\vp)  \right. \right. \nonumber\\
     \left. \left.
     -\left(v^2-\frac{7}{6}\frac{G\mc M}{r}\right)r\dot{\vp}\sin(\Phi-\vp) \right\} \right.
     \nonumber \\
     -\frac{1+\cos^2\Theta}{4}\left\{
     2\dot{r}\cos 3(\Phi-\vp)\left(\dot{r}^2-3r^2\dot{\vp}^2 -2\frac{G\mc M}{r}\right)
     \right. \nonumber \\
     \left. \left.
     -r\dot{\vp}\sin 3(\Phi-\vp)\left(7\frac{G\mc M}{r}-6\dot{r}^2+2r^2\dot{\vp}^2\right)
     \right\}
     \right], \\ 
\fl 
      \frac{Dc^5}{G}h_\times^{(1)}=\frac{\sin 2\Theta}{2}\frac{\Delta m}{\mc M}\mu \left[
      r\dot{\vp}\cos(\Phi-\vp)\left\{\frac{5}{2}\frac{G\mc M}{r}-v^2\right\}
       \right. \nonumber \\
     \left. 
      +\dot{r}\sin(\Phi-\vp)\left\{v^2-2\frac{G\mc M}{r}\right\}  
      +\frac{r\dot{\vp}}{2}\cos 3(\Phi-\vp)\left\{7\frac{G\mc M}{r}-6\dot{r}^2+2r^2\dot{\vp}^2\right\}
      \right. \nonumber \\
     \left.
      +\dot{r}\sin 3(\Phi-\vp)\left\{ \dot{r}^2-3r^2\dot{\vp}^2-2\frac{G\mc M}{r}\right\} 
      \right].
\end{eqnarray}
Note that these expressions depend on the mass difference and vanish for equal mass binaries. The next corrections to the gravitational waveforms read
\begin{eqnarray}
\fl 
    \frac{Dc^6}{G}h_+^{(2)}=\frac{1+\cos^2\Theta}{2}\mu\left[
    \cos 2(\Phi-\vp) \left\{
    (10-\nu)\frac{G^2\mc M^2}{r^2}+ \frac{9}{7}(1-3\nu)(\dot{r}^4-r^4\dot{\vp}^4)
    \right.
    \right. \nonumber \\
     \left. 
    +\frac{G\mc M}{r}\left(\frac{15+32\nu}{7}\dot{r}^2 
   - \frac{11+156\nu}{21} r^2\dot{\vp}^2
    \right) \right\}
    \nonumber \\
    \left. 
    +\frac{r\dot{r}\dot{\vp}}{7}\sin 2(\Phi-\vp)
     \left\{
    \frac{10}{3}(5+27\nu)\frac{G\mc M}{r}   
    + 18(1-3\nu)v^2\right\}
    \right] \nonumber \\
    +\frac{\sin^2\Theta}{2}\mu\left[
    (\nu-10)\frac{G^2\mc M^2}{r^2} -\frac{9}{7}(1-3\nu)v^4
    +\frac{G\mc M}{7r}\Bigl( (37-20\nu)r^2\dot{\vp}^2 
    \Bigr. \right. \nonumber \\
    \left. \Bigl.
    -(15+32\nu)\dot{r}^2 \Bigr) \right] \nonumber \\
    -\frac{1-3\nu}{56}\mu(7\cos^4\Theta-8\cos^2\Theta+1)\left[
    7\frac{G^2\mc M^2}{r^2}-\frac{G\mc M}{r}( 18\dot{r}^2+13r^2\dot{\vp}^2) +6v^4\right]
    \nonumber \\
    +\frac{1-3\nu}{42}\mu(7\cos^4\Theta-6\cos^2\Theta+1)\Bigl[
    \cos 2(\Phi-\vp) \Bigl\{ 6(r^4\dot{\vp}^4-\dot{r}^4) \Bigr.\Bigr. \nonumber \\
    \left. \left. 
     +\frac{G\mc M}{r}(18\dot{r}^2 -3r^2\dot{\vp}^2) 
    -7\frac{G^2\mc M^2}{r^2}\right\} 
    -r\dot{r}\dot{\vp}\sin 2(\Phi-\vp) \left\{ 12v^2 -27\frac{G\mc M}{r}\right\}
    \right]  \nonumber \\
    +\frac{1-3\nu}{24}\mu \sin^2\Theta(1+\cos^2\Theta)\left[
    \cos 4(\Phi-\vp) \left\{
    7\frac{G^2\mc M^2}{r^2}+6\dot{r}^4 -36r^2\dot{r}^2\dot{\vp}^2 
    \right. \right. \nonumber \\
    \left. \left. +6r^4\dot{\vp}^4 
    +\frac{G\mc M}{r}(51r^2\dot{\vp}^2-18\dot{r}^2) \right\}
   -r\dot{r}\dot{\vp}\sin 4(\Phi-\vp) \left\{
     54\frac{G\mc M}{r} -24 \dot{r}^2
    \right.\right. \nonumber \\
    \Bigl.\Bigl.
    +24r^2\dot{\vp}^2
    \Bigr\} 
    \Bigr] 
    -\frac{1-3\nu}{3} G\mc M\mu \dot{\vp}(2\cos^2\Theta-1) \Bigl[ 4r\dot{\vp}\cos 2(\Phi-\vp) \Bigr. \nonumber \\
    \Bigl. 
    -\dot{r}\sin 2(\Phi-\vp) \Bigr], 
\end{eqnarray}
\begin{eqnarray}
\fl 
      \frac{Dc^6}{G}h_\times^{(2)}=\mu \cos\Theta\left[
      \frac{r\dot{r}\dot{\vp}}{7}\cos 2(\Phi-\vp) \left\{ \frac{10}{3}(5+27\nu)\frac{G\mc M}{r}
      +18v^2(1-3\nu) \right\} 
      \right. \nonumber \\
      -\sin 2(\Phi-\vp) \left\{
      (10-\nu)\frac{G^2\mc M^2}{r^2} +\frac{G\mc M}{r} \left(
      \frac{15+32\nu}{7}\dot{r}^2 -\frac{11+156\nu}{21}r^2\dot{\vp}^2 \right)
      \right. \nonumber \\
      \left. \left.
      +\frac{9}{7}(1-3\nu)(\dot{r}^4-r^4\dot{\vp}^4) \right\} \right]
      +(1-3\nu)\mu\cos\Theta \left[
      -\frac{1}{2}G\mc M \dot{r}\dot{\vp}\sin^2\Theta 
      \right. \nonumber \\
     \left.
      +\frac{3\cos^2\Theta-1}{6} G\mc M\dot{\vp}\left\{ \dot{r}\cos 2(\Phi-\vp) \right\} +
      4r\dot{\vp}\sin 2(\Phi-\vp) \right\} \nonumber \\
      -\frac{\sin^2\Theta}{12} \left\{ r\dot{r}\dot{\vp}\cos 4(\Phi-\vp) \left(
      54\frac{G\mc M}{r} -24\dot{r}^2+24r^2\dot{\vp}^2\right) 
      \right. \nonumber \\
       \left.
      +\sin 4(\Phi-\vp) \left( 7\frac{G^2\mc M^2}{r^2}  
      +\frac{G\mc M}{r} (51r^2\dot{\vp}^2 -18\dot{r}^2) +6\dot{r}^4+6r^4\dot{\vp}^4 -36r^2\dot{r}^2\dot{\vp}^2 \right) \right\} \nonumber \\
      -\frac{7\cos^2\Theta-5}{42} \left\{ 3r\dot{r}\dot{\vp}\cos 2(\Phi-\vp) \left(4v^2-9\frac{G\mc M}{r}\right) 
     \right. \nonumber \\
      \left. \left. 
     + \sin 2(\Phi-\vp)\left( -7\frac{G^2\mc M^2}{r^2}  
     +\frac{G\mc M}{r}(18\dot{r}^2-3r^2\dot{\vp}^2)+6r^4\dot{\vp}^4-6\dot{r}^4\right)
     \right\}\right].
\end{eqnarray}
The polarization states of the gravitational radiation field, expressed in terms of generalized coordinates and momenta, can be found in appendix B.


\section{Discussion}
\label{sec:4}

\begin{figure}
 \begin{center}
  \fontsize{8}{12}
    \psfrag{t/T0}{$t/P_0$}
  \psfrag{hc}{$\hspace{-1cm} h_\times \ [G\mc M/Dc^2]$}  
  \psfrag{hp}{$\hspace{-1cm} h_+ \ [G\mc M/Dc^2]$}
  \psfrag{h0}{$h_+^{(0)}$}
  \psfrag{h2}{$h_+^{(1pN)}$}
  \resizebox{110mm}{!}{\includegraphics{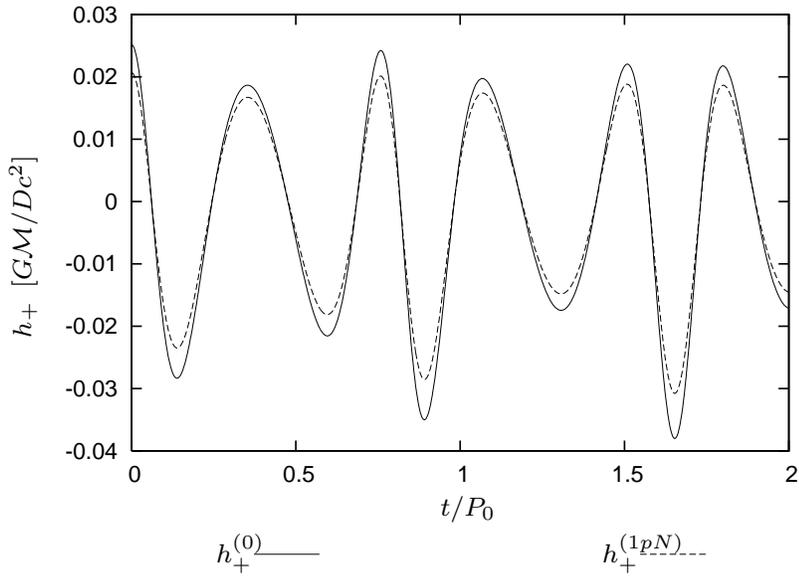}}
  \caption{$h_+$-component of a non-spinning, equal mass binary system with $\tilde{a}_r=40, e_r=0.3$.  Plotted are the waveform according to the leading order quadrupole formula and and the 1pN-corrected waveform. Observer-dependent parameters: $\Phi=\pi/2, \Theta=\pi/4$.}
\label{fig:1}
\end{center}
\end{figure}

\begin{figure}
 \begin{center}
  \fontsize{8}{12}
    \psfrag{t/T0}{$t/P_0$}
  \psfrag{hc}{$\hspace{-1cm} h_\times \ [G\mc M/Dc^2]$}  
  \psfrag{hp}{$\hspace{-1cm} h_+ \ [G\mc M/Dc^2]$}
  \psfrag{h0}{$h_\times^{(0)}$}
  \psfrag{h2}{$h_\times^{(1pN)}$}
  \resizebox{110mm}{!}{\includegraphics{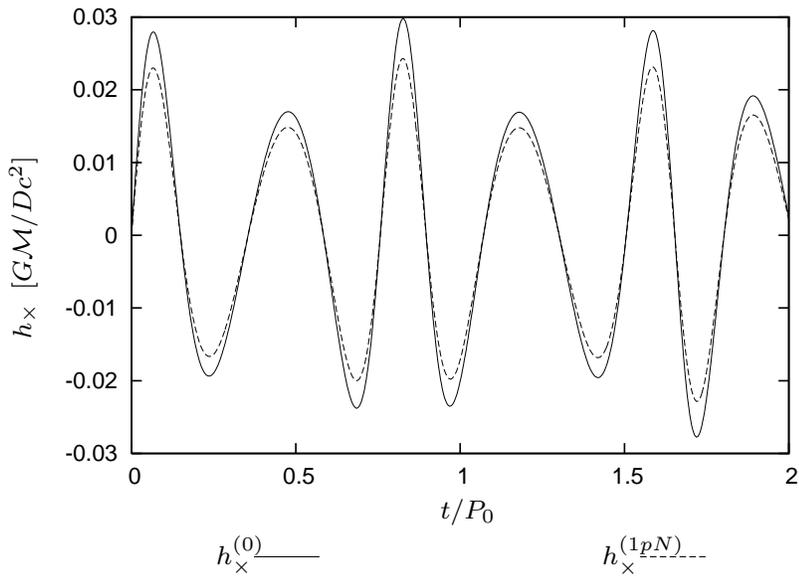}}
  \caption{$h_\times$-component of a non-spinning, equal mass binary system. All parameters are the same as in Fig. \ref{fig:1}. }
\label{fig:2}
\end{center}
\end{figure}

\begin{figure}
 \begin{center}
  \fontsize{8}{12}
  \psfrag{t/T0}{$t/P_0$}
  \psfrag{hp}{$\hspace{-1cm} h_+ \ [G\mc M/Dc^2]$}
  \psfrag{q=0}{$\hspace{-0.2cm} \tilde{q}=0$}
  \psfrag{q=4}{$\hspace{-0.2cm} \tilde{q}=4$}
  \includegraphics[width=110mm]{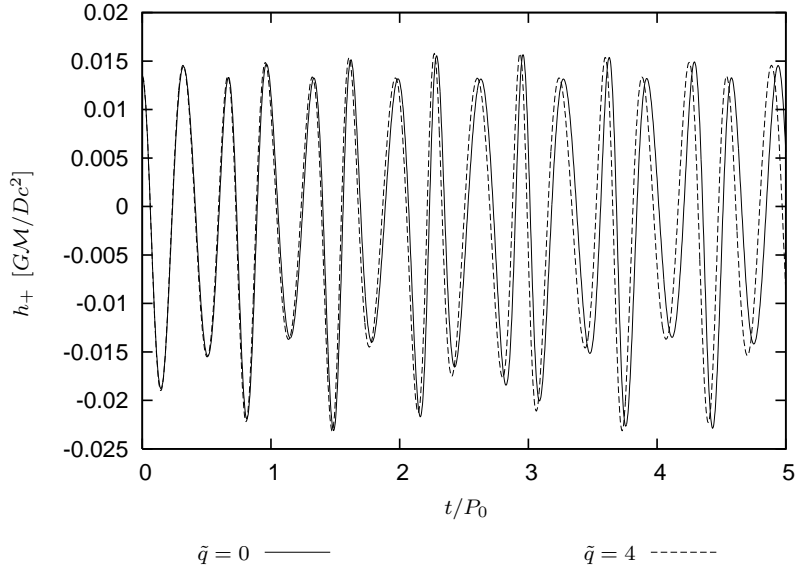}
  \caption{Influence of the quadrupole coupling on the gravitational wave emission. The 1pN correct $h_+$-component emitted by a non-spinning binary with semi-major axis and eccentricity $\tilde{a}_r=40$ and $e_r=0.3$, respectively, is compared with the corresponding waveform emitted by a NS-NS binary with $\tilde{q}=4$. The masses are $m_1=3m_2$. Observer-dependent angles are $\Phi=\pi/2, \Theta=\pi/4$.} 
 \label{fig:3}
 \end{center}
\end{figure}

\begin{figure}
 \begin{center}
  \fontsize{8}{12}
  \psfrag{t/T0}{$t/P_0$}
  \psfrag{hp}{$\hspace{-1cm} h_+ \ [G\mc M/Dc^2]$}
  \psfrag{q=0}{$\hspace{-0.2cm} \tilde{q}=0$}
  \psfrag{q=4}{$\hspace{-0.2cm} \tilde{q}=4$}
  \includegraphics[width=110mm]{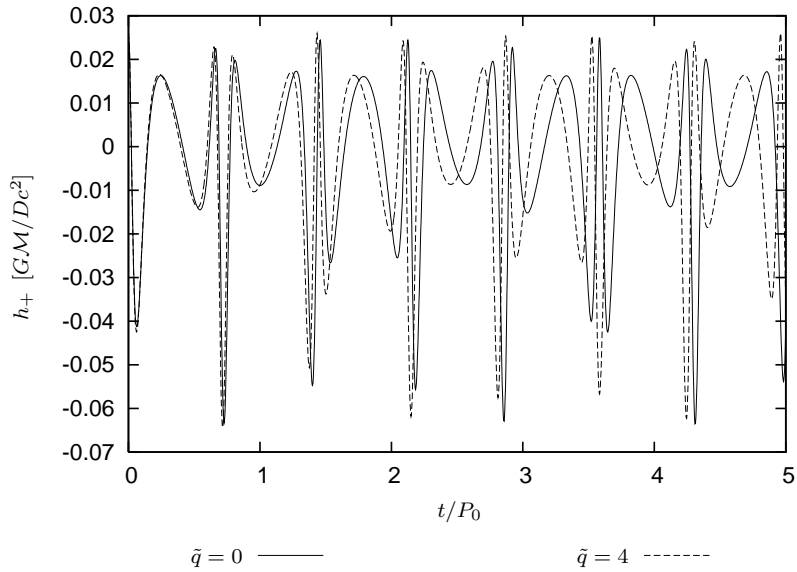}
  \caption{Influence of the quadrupole coupling on the gravitational wave emission. The 1pN correct $h_+$-component emitted by a non-spinning, equal mass  binary with semi-major axis and eccentricity $\tilde{a}_r=40$ and $e_r=0.6$, respectively, is compared with the corresponding waveform emitted by a NS-NS binary with $\tilde{q}=4$. Observer-dependent angles are $\Phi=\pi/2, \Theta=\pi/4$.} 
 \label{fig:4}
 \end{center}
\end{figure}

\begin{figure}
\fontsize{8}{12}
 \begin{center}
  \psfrag{hp}{$\hspace{-1cm} h_+\ [G\mc M/Dc^2]$}
  \psfrag{t/T0}{$t/P_0$}
  \resizebox{110mm}{!}{\includegraphics{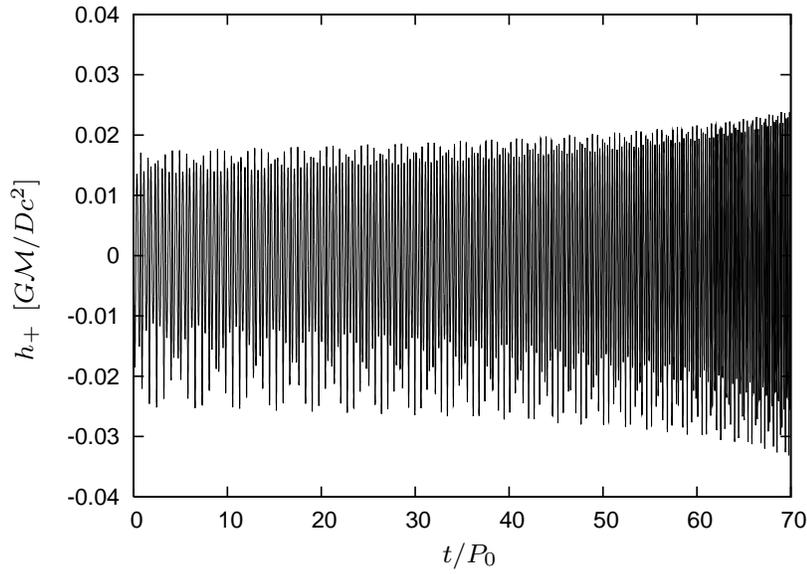}}
  \caption{$h_+$-component of the gravitational wave field emitted by a non-spinning, equal mass binary during the inspiral process (initial values $\td{a}_r(0)=50, e_r(0)=0.3$). Observer-dependent parameters $\Phi=\pi/2, \Theta=\pi/4$.}
  \label{fig:5}
 \end{center}
\end{figure}

\begin{figure}
\fontsize{8}{12}
 \begin{center}
  \psfrag{hp}{$\hspace{-1cm} h_+\ [G\mc M/Dc^2]$}
  \psfrag{t/T0}{$t/P_0$}
  \resizebox{110mm}{!}{\includegraphics{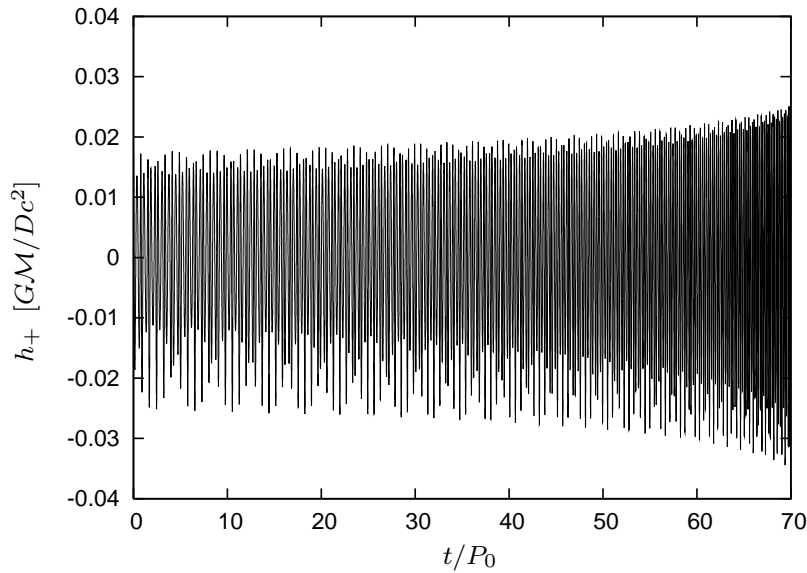}}
  \caption{$h_+$-component of the gravitational wave field emitted by an equal mass binary with $\td{q}=4$ during the inspiral process (initial values $\td{a}_r(0)=50, e_r(0)=0.3$). Observer-dependent parameters $\Phi=\pi/2, \Theta=\pi/4$.}
  \label{fig:6}
 \end{center}
\end{figure}

It has been long known that finize size effects introduce a periastron shift already at the level of Newtonian theory. For a couple of main sequence star binaries the total apsidal motion $\dot{\vp}_{tot}$ has been determined from observational evidence. Compared with the contribution $\dot{\vp}_{rel}$ predicted by GR it became obvious that in all systems the Newtonian perturbations give the dominant contribution to $\dot{\vp}_{tot}$ (for an overview see e.g. \cite{Claret2002}). This is due to the "soft" equations of state governing the stellar matter of main sequence stars. For compact star binaries  Newtonian perturbations are often neglected. In particular, it is often argued that the effect of the spin-induced quadrupole is too small unless the compact star (e.g. a neutron star) is rotating near the mass-shedding limit \cite{Wiseman1996}. However, even for NS-NS or NS-BH binaries 
this argument does not hold completely. It has been shown in previous sections that for close NS-NS binaries the potential energy introduced by the coupling can be considerably larger than the corresponding 1.5pN orbital correction terms, though it is smaller by a factor 100 or more than the 1pN contribution.  Thus, already at the level of the first post-Newtonian approximation in the EoM the rotational deformation induces a non-relativistic periastron shift, which accumulates over a large number of periods (figures  \ref{fig:3}) and \ref{fig:4}).
In order to obtain highly accurate templates it is thus desirable to take into account these corrections properly, at least for fast spinning neutron stars in a compact binary system.\\
With the space-bound laser interferometric detector LISA at hand the frequency band accessible to observations will be extended to much lower frequencies ($10^{-1}$ to $10^{-4}$Hz), which enlarges the number of possible sources enormeously. In particular, with LISA not only BH-BH, NS-NS and NS-BH binaries should be detectable, but also white dwarf binaries. In particular to this class of compact binaries the analysis shown in this paper applies. It has been argued by Willems \emph{et al.} in a recent paper \cite{Willems2007} that -- contrary to prevailing opinions -- there might exist a class of eccentric galactic double white dwarfs, which are formed by interactions in tidal clusters. Willems \emph{et al.} showed that tides and stellar rotation strongly dominate the periastron shift at orbital frequencies $\ge$1 mHz. The phase shifts induced by these Newtonian perturbations are much larger than the general relativistic corrections then. They conclude that it is essential to include phase shifts generate by Newtonian perturbations into the signal templates in order to not bias LISA surveys against eccentric double white dwarfs. Generally, neglecting the contribution to $\dot{\vp}_{tot}$ induced by rotational deformation will lead to an overestimation of the total mass derived from $\dot{\vp}_{tot}$. \\ 
In this paper the competing influences of the rotational deformation and the 1pN correction terms were examined in more detail. In particular, we succeeded in calculating a  1pN quasi-Keplerian solution, which takes into account finite size effects up to linear order in the quadrupole deformation parameter $q$. The results given in section 2 are valid as long as $q/J^2$ is of the order $O(c^{-2})$. For white dwarf binaries or binary pulsars such as PSR 1259-63 the periastron shift induced by rotational deformation is possibly much larger than the general relativistic contribution. In that case, Eqs. (\ref{sol}-\ref{sol2}) still apply in the limit $v/c\to 0$. In section 3 the polarization states of the gravitational radiation field are calculated beyond the leading order approximation. For non-spinning compact binaries the corresponding waveforms are shown in figures \ref{fig:1} and \ref{fig:2}. In these figures waveforms calculated using the leading order expressions $h_+,\times^{(0)}$ are compared to the 1pN correct waveforms with the next to leading order corrections $h_{+,\times}^{(1)}$ and $h_{+,\times}^{(2)}$ taken into account. The first correction, $h_{+,\times}^{(1)}$, is nontrivial only for different mass binaries, i.e. for equal-mass binaries the first non-vanishing correction to the leading order formula appears at the order $O(c^{-2})$.
\\
The influence of the $q$-coupling on the gravitational waveforms is shown in figures \ref{fig:3} and \ref{fig:4}. As expected, the spin-induced quadrupole moment leads to a phase shift compared to the pure point-particle GW emission. Moreover, the quadrupole deformation of the spinning compact objects speeds up the inspiral process, as has been shown in figures \ref{fig:5} and \ref{fig:6} for an equal-mass binary in a slightly elliptic orbit. 
\\
More analysis is needed in order to fully understand the imprint of finite size effects onto the gravitational wave pattern of close compact binary systems beyond the leading order. In particular it would be highly desirable to include the stellar oscillation modes into the calculations. From previous works  it is expected that in these cases so called tidal resonances will have an important impact on the inspiral process and the gravitational wave emission of the binary  \cite{Kokkotas1995}, \cite{Ho1999}, \cite{Hansen2005}.

\subsection*{Acknowledgements}
 I am grateful to Gerhard Sch\"afer for helpful discussions and careful reading of the manuscript. This work is supported by the Deutsche Forschungsgemeinschaft (DFG) through SFB/TR7 "Gravitationswellenastronomie".


\begin{appendix}

\section{Useful relations}

The mass and current multipole moments  $I^{lm}$ and $S^{lm}\ (m=-l,\dots,l)$ that are irreducibly defined with respect to the orbital angular momentum axis are related to $\mc I_{A_l}$ and $\mc J_{A_l}$ according to
\begin{eqnarray}
\label{rel1}
    I^{lm}(t)&=\frac{16\pi}{(2l+1)!!}\sqrt{\frac{(l+1)(l+2)}{2(l-1)l}}\mc I_{A_l}(t)Y_{A_l}^{lm\ast},\\
\label{rel2}
   S^{lm}(t)&=-\frac{32\pi l}{(l+1)(2l+1)!!}
   \sqrt{\frac{(l+1)(l+2)}{2(l-1)l}}\mc J_{A_l}Y_{A_l}^{lm\ast},
\end{eqnarray}
where, for $m\ge 0$,
\begin{eqnarray}
\fl
    Y_{A_l}^{lm}=(-1)^m(2l-1)!!\sqrt{\frac{2l+1}{4\pi(l-m)!(l+m)!}}
    (\delta_{\lan i_1}^1+i\delta_{\lan i_1}^2)
    \cdots(\delta_{i_m}^1+i\delta_{i_m}^2)\delta_{i_{m+1}}^3\cdots\delta_{i_l\ran}^3,
\end{eqnarray}
and
\begin{eqnarray}
     Y^{lm}_{A_l}=(-1)^mY^{l|m|\ast}_{A_l}\qquad
     \mathrm{for}\qquad m<0.
\end{eqnarray}
The complex conjugates are given by 
\begin{eqnarray}
      I^{lm\ast}=(-1)^mI^{l-m},\qquad
      S^{lm\ast}=(-1)^mS^{l-m}.
\end{eqnarray}
The pure-spin tensor-spherical harmonics are orthonormal on the unit sphere. For the complex conjugate the following relation holds:
\begin{eqnarray}
     T^{E/B2,lm\ast}=(-1)^mT^{E/B2,l-m}.
\end{eqnarray}
Defining
\begin{eqnarray*}
    \vec\Upsilon_+\equiv \hat{\Theta}\otimes\hat{\Theta}-\hat{\Phi}\otimes\hat{\Phi},\qquad
    \vec\Upsilon_-\equiv \hat{\Theta}\otimes\hat{\Phi}+\hat{\Phi}\otimes\hat{\Theta}
\end{eqnarray*}
the expressions needed in the paper read
\begin{eqnarray}
\fl 
\label{pure1}
     T^{E2,22}=\sqrt{\frac{5}{128\pi}}e^{2i\Phi} \Bigl[ (1+\cos^2\Theta)\vec\Upsilon_+ +2i\cos\Theta \vec\Upsilon_-\Bigr], \\
\fl 
\label{pure2}
     T^{E2,20}=\sqrt{\frac{15}{64\pi}}\sin^2\Theta \vec\Upsilon_+, \\
\fl 
\label{pure3}
     T^{B2,21}=-\sqrt{\frac{5}{32\pi}}\sin\Theta e^{i\Phi}\Bigl[ i\vec\Upsilon_+ -\cos\Theta \vec\Upsilon_-\Bigr], \\
\fl 
\label{pure4}
     T^{E2,33}=-\sqrt{\frac{21}{256\pi}} \sin\Theta e^{3i\Phi} \Bigl[
     (1+\cos^2\Theta)\vec\Upsilon_+
     +2i\cos\Theta\vec\Upsilon_-\Bigr], \\
\fl 
\label{pure5}
     T^{E2,31}=\sqrt{\frac{35}{256\pi}} \sin\Theta e^{i\Phi}\Bigl[
     (3\cos^2\Theta-1)\vec\Upsilon_+ +2i\cos\Theta\vec\Upsilon_- \Bigr],
\\
\fl 
\label{pure6}
    T^{B2,32}=-\sqrt{\frac{7}{128\pi}}e^{2i\Phi}\Bigl[ 2i(2\cos^2\Theta-1)\vec\Upsilon_+ 
    -\cos\Theta
    (3\cos^2\Theta-1)\vec\Upsilon_-\Bigr],\\
\fl 
\label{pure7}
     T^{B2,30}=\sqrt{\frac{105}{64\pi}}\cos\Theta\sin^2\Theta\vec\Upsilon_-. \\
\fl 
\label{pure8}
     T^{E2,44}=\sqrt{\frac{63}{512\pi}}\sin^2\Theta e^{4i\Phi}\Bigl[
     (1+\cos^2\Theta)\vec\Upsilon_+
     +2i\cos\Theta\vec\Upsilon_-\Bigr], \\
\fl 
\label{pure9}
     T^{E2,42}=\sqrt{\frac{9}{128\pi}}e^{2i\Phi} \Bigl[
     (7\cos^4\Theta-6\cos^2\Theta+1)\vec\Upsilon_+
     +i\cos\Theta(7\cos^2\Theta-5)\vec\Upsilon_-\Bigr],\\
\fl
\label{pure10}
    T^{E2,40}=-\sqrt{\frac{45}{256\pi}}(7\cos^4\Theta-8\cos^2\Theta+1)\vec\Upsilon_+.
\end{eqnarray}

\subsection{Symmetric tracefree tensors}

Throughout this paper  symmetric-tracefree 3$rd$ and 4$th$ rank tensors  are used. Symmetrizing a tensor of rank $p$ requires to take the properly weighted sum over all index permutations,
\begin{eqnarray}
     T_{(i_1\dots i_p)}=
     T_{i_1\dots i_p}^{symm}\equiv \frac{1}{p!}\sum_{permut.}T_{i_1\dots i_p}.
\end{eqnarray}
The tracefree part of the tensor $T_{i_1\dots i_p}$  is calculated according to \cite{Pirani1965}
\begin{eqnarray}
  T_{\lan i_1\dots i_p\ran}=\sum_{k=0}^{[p/2]} a_k^p\delta_{(i_1i_2}\cdots
  \delta_{i_{2k-1}i_{2k}}T^{symm}_{i_{2k+1}\dots i_p)\alpha_1\alpha_1\dots \alpha_k\alpha_k},
\end{eqnarray}
with
\begin{eqnarray}
    a_k^p=\frac{p!}{(2p-1)!!}\frac{(-1)^k(2p-2k-1)!!}{(p-2k)!(2k)!!}.
\end{eqnarray}
In particular, 3$rd$ and 4$th$ rank STF tensors are given by
\begin{eqnarray}
\fl 
    T_{\lan abc\ran}=T_{(abc)}-\frac{1}{5}\left[
    \delta_{ab}T_{(cii)}+\delta_{bc}T_{(aii)}+\delta_{ac}T_{(bii)}\right], \\
\fl
    T_{\lan abcd\ran}=
     T_{(abcd)}-\frac{1}{7}[\delta_{ab}T_{(cdii)}+\delta_{ac}T_{(bdii)}+\delta_{ad}T_{(bcii)} 
   +\delta_{bc}T_{(adii)}+\delta_{bd}T_{(acii)}
   +\delta_{cd}T_{(abii)}] \nonumber \\
    +\frac{1}{35}[\delta_{ac}\delta_{bd}+\delta_{ad}\delta_{bc}+\delta_{ab}\delta_{cd}]T_{(iijj)}.
\end{eqnarray}

 \section{Expressions for $h_+$ and $h_\times$ in terms of generalized coordinates and momenta}

The expressions for the leading and next to leading order contribution to the polarization states of the gravitational wave field, $h_+$ and $h_\times$, read
\begin{eqnarray}
\fl 
   h_+^{(0)}=\frac{G\mu}{Dc^4} \left\{
  \frac{1+\cos^2\Theta}{\mu^2}\left[ \cos2(\Phi-\vp)\left\{ p_r^2-\frac{p_\vp^2}{r^2}-\frac{G\mc M\mu^2}{r}\left(1+\frac{3q}{2r^2}\right)\right\} \right.  \right. \nonumber \\
    \left. \left. 
   +2\frac{p_rp_\vp}{r}\sin2(\Phi-\vp)\right] 
  -\frac{\sin^2\Theta}{\mu^2}\left[ p_r^2+\frac{p_\vp^2}{r^2}-\frac{G\mc M\mu^2}{r}\left(1+\frac{3q}{2r^2}\right)\right]
   \right\}, \\
\fl 
   h_+^{(1)}=\frac{G}{Dc^5}\frac{\Delta m}{\mc M\mu^2}\sin\Theta\left[
   \frac{p_\vp}{r}\sin(\Phi-\vp)\left\{
   \frac{4}{3}\frac{G\mc M\mu^2}{r} 
   \right.\right. \nonumber \\
   \left. \left. 
    +\frac{3\cos^2\Theta-1}{2}\left(\frac{7}{6}\frac{G\mc M\mu^2}{r}-p_r^2-\frac{p_\vp^2}{r^2}\right)\right\} 
   \right.\nonumber \\
   +\frac{3\cos^2\Theta-1}{2}p_r\cos (\Phi-\vp)\left\{ 2\frac{G\mc M\mu^2}{r} -p_r^2-\frac{p_\vp^2}{r^2} \right\} \nonumber \\
   -\frac{1+\cos^2\Theta}{2}\left\{ p_r\cos 3(\Phi-\vp)\left(
   p_r^2-3\frac{p_\vp^2}{r^2} -2\frac{G\mc M\mu^2}{r}\right) \right. \nonumber \\
   \left. \left. 
   -\frac{p_\vp}{r} \sin 3(\Phi-\vp) \left(\frac{7}{2}\frac{G\mc M\mu^2}{r} -3p_r^2 +\frac{p_\vp^2}{r^2} \right) \right\} \right],
\end{eqnarray}
\begin{eqnarray}
\fl 
    h^{(2)}_+= \frac{G\mu}{Dc^6}\left\{
    \frac{\sin^2\Theta}{14} \left[ 7\frac{G^2\mc M^2}{r^2}(\nu-10)
   -\frac{5(3\nu-1)}{\mu^4}\left( p_r^2+\frac{p_\vp^2}{r^2}\right)^2
   \right. \right. \nonumber \\
   \left. \left. 
   +\frac{G\mc M}{\mu^2 r}\left\{3(23+8\nu)p_r^2+(121+8\nu)\frac{p_\vp^2}{r^2}\right\}
   \right] \right. \nonumber \\
   +\frac{1+\cos^2\Theta}{14}\cos2(\Phi-\vp)\left[
    7\frac{G^2\mc M^2}{r^2}(10-\nu)
   +5(3\nu-1)\left(\frac{p_r^4}{\mu^4}-\frac{p_\vp^4}{\mu^4r^4}\right) \right. \nonumber \\
   \left.
   +\frac{G\mc M}{\mu^2 r}\left\{\frac{241-72\nu}{3}\frac{p_\vp^2}{r^2}-3(23+8\nu)p_r^2\right\}
  \right] \nonumber \\
   +\frac{1+\cos^2\Theta}{7}\sin2(\Phi-\vp)\frac{p_rp_\vp}{\mu^2 r}\left[
     \frac{5(3\nu-1)}{\mu^2}\left(p_r^2+\frac{p_\vp^2}{r^2}\right) -\frac{227-9\nu}{3}\frac{G\mc M}{r}\right] \nonumber \\
   +\frac{1-3\nu}{24}\sin^2\Theta(1+\cos^2\Theta)\left[ \cos4(\Phi-\vp)\left\{
   7\frac{G^2\mc M^2}{r^2}-\frac{G\mc M}{\mu^2r}\left(18p_r^2-51\frac{p_\vp^2}{r^2}\right) \right.\right. \nonumber \\
   \left. \left.
    +6\left(\frac{p_r^4}{\mu^4}-6\frac{p_r^2p_\vp^2}{\mu^4r^2}+\frac{p_\vp^4}{\mu^4r^4}
   \right)\right\} 
   -\frac{p_rp_\vp}{\mu^2r}\sin4(\Phi-\vp)\left\{54\frac{G\mc M}{r}   \right.\right. \nonumber \\
    \left.\left. 
    +24\left(\frac{p_\vp^2}{\mu^2r^2}-\frac{p_r^2}{\mu^2}\right)\right\}\right]
   +\frac{1-3\nu}{42} (7\cos^4\Theta-6\cos^2\Theta+1)\Bigl[ \Bigr.  \nonumber \\
   \cos2(\Phi-\vp)\left\{
     -7\frac{G^2\mc M^2}{r^2}+\frac{G\mc M}{\mu^2r}\left(18p_r^2-3\frac{p_\vp^2}{r^2}\right) +6\left(\frac{p_\vp^4}{\mu^4r^4}-\frac{p_r^4}{\mu^4}\right)\right\}  \nonumber 
   \\
   \left.
   -\frac{p_rp_\vp}{\mu^2r}\sin2(\Phi-\vp)\left\{ 12\left(\frac{p_r^2}{\mu^2}+\frac{p_\vp^2}{\mu^2r^2}\right)-27\frac{G\mc M}{r}\right\}
    \right] \nonumber \\
  -\frac{1-3\nu}{56} (7\cos^4\Theta-8\cos^2\Theta+1)\left[7\frac{G^2\mc M^2}{r^2}
  +6\left(\frac{p_r^2}{\mu^2}+\frac{p_\vp^2}{\mu^2r^2}\right)^2
  \right. \nonumber \\
  \left. 
   -\frac{G\mc M}{\mu^2r}\left(18p_r^2+13\frac{p_\vp^2}{r^2}\right)\right] \nonumber \\
   \left.
   -\frac{1-3\nu}{3}(2\cos^2\Theta-1)\frac{G\mc Mp_\vp}{\mu^2 r^2}\left[4\frac{p_\vp}{r}\cos2(\Phi-\vp)-
   p_r\sin2(\Phi-\vp)\right]   \right\}.
\end{eqnarray}
For $h_\times$ one finds
\begin{eqnarray}
\fl 
 h_\times^{(0)}=2\frac{G\mu}{Dc^4} \cos\Theta\left[\sin2(\Phi-\vp)\left\{-\frac{p_r^2}{\mu^2}+\frac{p_\vp^2}{\mu^2r^2}+\frac{G\mc M}{r}\left(1+\frac{3q}{2r^2}\right)\right\} \right. \nonumber \\
   \left. 
   +2\frac{p_rp_\vp}{\mu^2r}\cos2(\Phi-\vp)\right],\\
\fl 
  h_\times^{(1)}=\frac{G}{Dc^5}\frac{\Delta m}{\mc M\mu^2}\frac{\sin 2\Theta}{2} \left[
    \frac{p_\vp}{r}\cos (\Phi-\vp) \left\{ \frac{5}{2}\frac{G\mc M\mu^2}{r} -p_r^2-\frac{p_\vp^2}{r^2} \right\} \right. \nonumber \\
     +p_r\sin(\Phi-\vp)\left\{ p_r^2+\frac{p_\vp^2}{r^2} -2\frac{G\mc M\mu^2}{r} \right\}
     +\frac{p_\vp}{r}\cos 3(\Phi-\vp) \left\{ 7\frac{G\mc M\mu^2}{r} 
     \right. \nonumber \\
     \left. 
     \left. 
     -6p_r^2+2\frac{p_\vp^2}{r^2}
     \right\} 
     +p_r\sin 3(\Phi-\vp)\left\{ p_r^2-3\frac{p_\vp^2}{r^2}-2\frac{G\mc M\mu^2}{r}\right\}
     \right],
\end{eqnarray}
\begin{eqnarray}
\fl 
 h^{(2)}_{\times}=\frac{G\mu}{Dc^6}\cos\Theta\left[
   \sin2(\Phi-\vp)\left\{\frac{5}{7}\frac{(1-3\nu)}{\mu^4}\left(p_r^4-\frac{p_\vp^4}{r^4}\right)+
   \frac{G\mc M}{\mu^2r}\left(\frac{3}{7}(23+8\nu)p_r^2
    \right.\right.\right. \nonumber \\ 
   \left.\left.\left. 
   -\frac{241-72\nu}{21}\frac{p_\vp^2}{r^2}\right) 
   -\frac{G^2\mc M^2}{r^2}(10-\nu)\right\}   \right.  \\
  +\frac{p_rp_\vp}{\mu^2r}\cos2(\Phi-\vp)\left\{\frac{10}{7}\frac{(3\nu-1)}{\mu^2}\left(p_r^2+\frac{p_\vp^2}{r^2}\right)
  -\frac{2}{21}(227-9\nu)\frac{G\mc M}{r}\right\} \nonumber \\
  -\frac{1-3\nu}{12}\frac{p_rp_\vp}{\mu^2r}\sin^2\Theta\cos4(\Phi-\vp)\left\{54\frac{G\mc M}{r} +24\left(\frac{p_\vp^2}{\mu^2r^2}-\frac{p_r^2}{\mu^2}\right)\right\} \nonumber \\
    -\frac{1-3\nu}{12}\sin^2\Theta\sin4(\Phi-\vp)\left\{
    7\frac{G^2\mc M^2}{r^2} +\frac{G\mc M}{\mu^2r}
   \left(51\frac{p_\vp^2}{r^2}-18p_r^2\right)
    \right. \nonumber \\
   \left. 
   +6\left(\frac{p_r^4}{\mu^4}-6\frac{p_r^2p_\vp^2}{\mu^4r^2}+\frac{p_\vp^4}{\mu^4r^4}\right)
    \right\}
    -\frac{1-3\nu}{2}\sin^2\Theta\frac{G\mc M}{r}\frac{p_rp_\vp}{\mu^2r}   \nonumber \\
   -\frac{1-3\nu}{42}(7\cos^2\Theta-5)\left[\frac{p_rp_\vp}{\mu^2 r}\cos2(\Phi-\vp)\left\{
   12\left(\frac{p_r^2}{\mu^2}+\frac{p_\vp^2}{\mu^2r^2}\right) 
   -27\frac{G\mc M}{r}\right\} 
   \right.\nonumber \\
    \left.+
    \sin2(\Phi-\vp)\left\{
    -7\frac{G^2\mc M^2}{r^2}+\frac{G\mc M}{\mu^2r}\left(18p_r^2-3\frac{p_\vp^2}{r^2}\right) +6\left(\frac{p_\vp^4}{\mu^4r^4}-\frac{p_r^4}{\mu^4}\right)
    \right\}
    \right] \nonumber \\
    \left.
    +\frac{1-3\nu}{6}(3\cos^2\Theta-1)\frac{G\mc M}{r}
    \left\{\frac{p_rp_\vp}{\mu^2r}\cos2(\Phi-\vp)+4\frac{p_\vp^2}{\mu^2r^2}\sin2(\Phi-\vp)\right\}
    \right] \nonumber.
\end{eqnarray}

\section{Higher order gravitational wave forms: Analytic expressions}

Using the quasi-Keplerian parametrization derived in section 2 it is possible to calculate analytic expressions for the times derivatives of the STF multipole moments entering in the multipole expansion (\ref{rad2}). Defining $F(u)\equiv 1-e_r\cos u$ one obtains
\begin{eqnarray}
 S^{(2)21}&=\frac{32}{3}\sqrt{\frac{\pi}{5}}(m_1-m_2)\nu(-E)^{3/2} \frac{
 \sqrt{1-e_r^2}}{F(u)^2},\\
 S^{(3)30}&=-16\mu E^2\sqrt{\frac{\pi}{105}}(1-3\nu)\frac{e_r\sin u\sqrt{1-e_r^2}}{F(u)^3},\\
 S^{(3)32}&=\frac{8}{3}\sqrt{\frac{2\pi}{7}}(1-3\nu)\mu E^2 e^{-2i\vp} \frac{\sqrt{1-e_r^2}}{F(u)^3}
 \left[ e_r\sin u -4i\sqrt{1-e_r^2}\right],
\end{eqnarray}
while the time derivatives of the mass multipole moments read
\begin{eqnarray}
\fl 
 I^{(2)20}=-16\mu E\sqrt{\frac{\pi}{15}} \left[
 1-\frac{1}{F(u)}\left\{1-\frac{q}{2a_r^2F(u)^2}\right\} +\frac{E}{F(u)}\delta +\frac{E}{14c^2}
 \Bigl\{
  3(3\nu-1) \Bigr.\right. \nonumber \\
  \left.\left. 
  -\frac{51\nu -115}{F(u)}  
   +\frac{2(19\nu-4)}{F(u)^2} +4(\nu-26)\frac{1-e_r^2}{F(u)^3} \right\} \right],\\
\fl 
 I^{(2)22}= 4\sqrt{\frac{8\pi}{5}} \mu E e^{-2i\vp} \Bigl[
  -1+\frac{3}{F(u)}-\frac{2e_r^2\sin^2 u}{F(u)^2} +2i\frac{e_r\sqrt{1-e_r^2}\sin u}{F(u)^2}
   +\frac{5q}{2a_r^2 F(u)^3} \Bigr. \nonumber \\
   -E\delta \left\{ \frac{3}{F(u)}+\frac{4e_r^2\sin^2u}{F(u)^3} +
    2i\frac{e_r\sqrt{1-e_r^2}\sin u}{F(u)^2} \left( \frac{1+e_r^2}{1-e_r^2} -\frac{e_r\cos u}{F(u)}
   \right) \right\} \nonumber \\ 
   +\frac{E}{42 c^2} \Bigl\{
   9(3\nu-1) -\frac{3(51\nu-115)}{F(u)} +\frac{42(8\nu-25) -18e_r^2(3\nu-1)}{F(u)^2} 
    \Bigr. \nonumber \\
   \left. \left.  
    -4(111\nu - 254) \frac{1-e_r^2}{F(u)^3} 
   +\frac{2i e_r\sin u}{\sqrt{1-e_r^2}F(u)^3} \left(
   253-171\nu -3(23\nu-87)e_r\cos u
   \right.\right.\right. \nonumber \\
   \Bigl.\Bigl.\left.
   +(213\nu -505)e_r^2+9(3\nu-1)e_r^3\cos u\right) \Bigr\} 
   \Bigr], \\
\fl 
   I^{(3)31}=8\sqrt{\frac{2\pi}{35}}(m_1-m_2)\nu(-E)^{3/2}e^{-i\vp} \left[\frac{e_r\sin u}{F(u)}
   -i\frac{\sqrt{1-e_r^2}}{F(u)}\left( 1-\frac{5/6}{F(u)}\right)\right], \\
\fl 
   I^{(3)33}=8\sqrt{\frac{2\pi}{21}}(m_1-m_2)\nu (-E)^{3/2} e^{-3i\vp} \left[
   -\frac{e_r\sin u}{F(u)} \left\{ 1+\frac{4(1-e_r^2)}{F(u)^2}\right\} \right. 
   \nonumber \\
    \left.
   +i\frac{\sqrt{1-e_r^2}}{F(u)} \left\{ 3-\frac{5/2}{F(u)} +\frac{4(1-e_r^2)}{F(u)^2} \right\}
    \right],\\
\fl 
   I^{(4)40}=\frac{8}{21}\sqrt{\frac{\pi}{5}}(1-3\nu)\mu E^2 \left[ 6-\frac{6}{F(u)}-\frac{5}{F(u)^2} +\frac{5(1-e_r^2)}{F(u)^3} \right], \\
\fl 
   I^{(4)42}=\frac{8}{63}\sqrt{2\pi}(1-3\nu) \mu E^2 e^{-2i\vp} \left[ -6+\frac{6}{F(u)} 
   -\frac{7-12e_r^2}{F(u)^2} +\frac{3(1-e_r^2)}{F(u)^3} \right. \nonumber \\
     \left. 
    -3i\frac{e_r\sqrt{1-e_r^2}\sin u}{F(u)^2} \left\{4+\frac{1}{F(u)}\right\} \right], \\
\fl
   I^{(4)44}=\frac{4}{9}\sqrt{\frac{2\pi}{7}}(1-3\nu)\mu E^2 e^{-4i\vp} \left[
    6-\frac{6}{F(u)} +\frac{43-48e_r^2}{F(u)^2} -\frac{27(1-e_r^2)}{F(u)^3}
    +\frac{48(1-e_r^2)^2}{F(u)^4} \right. \nonumber \\
   \left. 
   +6i\frac{e_r\sqrt{1-e_r^2}\sin u}{F(u)^2} \left\{ 4+\frac{1}{F(u)}+\frac{8(1-e_r^2)}{F(u)^2} 
   \right\}\right].
\end{eqnarray}

\end{appendix}

\newpage

\end{document}